\documentclass[preprint,aip,jcp]{revtex4-1}
\usepackage{epsfig}
\usepackage{textcomp}
\usepackage{graphicx}
\usepackage{amsmath}
\usepackage{amssymb}
\usepackage{amsmath}
\usepackage{url}
\usepackage[normalem]{ulem}

\usepackage{color}
\definecolor{Myblue}{rgb}{0.3,0.3,1.0}

\usepackage{xr}
\newcommand*{\citen}[1]{%
  \begingroup
    \romannumeral-`\x 
    \setcitestyle{numbers}%
    \cite{#1}%
  \endgroup   
}

\usepackage{rotating}

\begin{document}
\title{Accurate Reproducing Kernel-Based Potential Energy Surfaces for
  the Triplet Ground States of N$_2$O and Dynamics for the
  N+NO$\leftrightarrow$O+N$_2$ and N$_2$+O$\rightarrow$2N+O Reactions}
  
\author{Debasish Koner} \affiliation{Department of Chemistry,
  University of Basel, Klingelbergstrasse 80, CH-4056 Basel,
  Switzerland} 
  
\author{Juan Carlos San Vicente Veliz} \affiliation{Department of Chemistry,
  University of Basel, Klingelbergstrasse 80, CH-4056 Basel,
  Switzerland}

\author{Raymond J. Bemish} \affiliation{Air Force Research Laboratory,
  Space Vehicles Directorate, Kirtland AFB, New Mexico 87117, USA}

\author{Markus Meuwly} \email[]{m.meuwly@unibas.ch}
\affiliation{Department of Chemistry, University of Basel,
  Klingelbergstrasse 80, CH-4056 Basel, Switzerland}
  
\date{\today}

\begin{abstract}
Accurate potential energy surfaces (PESs) have been determined for the
$^3$A$'$ and $^3$A$''$ states of N$_2$O using electronic structure
calculations at the multireference configuration interaction level
with Davidson correction (MRCI+Q) and the augmented Dunning-type
correlation consistent polarize triple zeta (aug-cc-pVTZ) basis
set. More than 20000 MRCI+Q/aug-cc-pVTZ energies are represented using a
reproducing kernel Hilbert space (RKHS) interpolation scheme. The RKHS
PESs successfully describe all reactant channels with high
accuracy. The analytical PESs are characterized by computing the
minima and transition states on it. Quasiclassical dynamics
simulations are then used to determine thermal and vibrational
relaxation rates for the N+NO and O+N$_2$ collisions. The agreement
between results obtained from the simulations and from available
experiments is favourable for both types of observables, which
provides a test for the accuracy of the PESs. The PESs can be used to
calculate more detailed state-to-state observables relevant for
applications to hypersonic reentry.
\end{abstract}
 
\maketitle

\section{Introduction}
The [NNO] reactive collision system is relevant for atmospheric and
environmental chemistry, and at the hypersonic flight
regime.\cite{sto71:8384,bra73:939,cap11:124007} Two particularly
relevant reactions involve the N+NO $\leftrightarrow$ O+N$_2$
processes. Both the forward (N+NO $\rightarrow$ O+N$_2$) and reverse
(O+N$_2$ $\rightarrow$ N+NO) reactions are major processes in
modelling the hyperthermal interactions during reentry of space
vehicles and thus play significant role in aerospace engineering. The
forward reaction produces molecular nitrogen while the reverse
reaction generates nitric oxide via the Zeldovich
mechanism.\cite{zeldovich:1946} The forward reaction also plays
important role in the Martian and Venusian
atmospheres.\cite{sis92:3209} The ground state of the N$_2$O molecule
is singlet ($^1$A$'$) and asymptotically connects with the
N($^1$D)+NO(X$^2\Pi)$ and O($^1$D)+N$_2$(X$^1\Sigma_g^+$)
channels. For the singlet electronic state of N$_2$O the
photodissociation dynamics and reactive collisions have been studied
previously using theory and
experiment.\cite{bla69:116,fel90:4768,sch10:091103,sch12:142,li12:4646,li14:1277}
On the other hand, the triplet manifold of N$_2$O connects with the
N($^4$S)+NO(X$^2\Pi)$ and O($^3$P)+N$_2$(X$^1\Sigma_g^+$) states, with
all the atomic and diatomic fragments in their electronic ground
states in both channels.  The N($^4$S)+NO(X$^2\Pi)$ $\rightarrow$
O($^3$P)+N$_2$(X$^1\Sigma_g^+$) is highly exothermic and plays an
important role in the upper atmosphere in removing
N($^4$S).\cite{war88} The reaction also helps removing the NO
pollutant from the atmosphere which leads to ozone depletion.  The
$^4$S state of N and $^2\Pi$ state of NO lead to $^3$A$'$, $^3$A$''$,
$^5$A$'$ and $^5$A$''$ states in $C_s$ symmetry while the
O($^3$P)+N$_2$($^1\Sigma_g^+$) channel results $^3$A$'$ and $^3$A$''$
states.\cite{wal87:6946} In the absence of spin orbit coupling both
reactions occurs adiabatically on the triplet PESs.\\

\noindent
For the forward reaction, several experiments have reported rates
using different techniques. One experiment measured the N-atom
concentration via line absorption and reported a rate of $3 \pm 1
\times 10^{-11}$cm$^3$s$^{-1}$ at 300 K.\cite{lin70:3896} Conversely,
using discharge flow-resonance fluorescence (DF-RF) and flash
photolysis-resonance fluorescence and a value of $3.4 \pm 0.9 \times
10^{-11}$cm$^3$s$^{-1}$molecule$^{-1}$ was suggested in the 196--400 K
temperature range.\cite{lee78:3069} This rate was recommended for
chemical modeling of stratospheric processes,\cite{dem83} while
another DF-RF experiment\cite{wen94:18839} reported a rate of
$(2.2\pm0.2)\times 10^{-11}$
exp$(160\pm50/T)$cm$^3$s$^{-1}$molecule$^{-1}$ over the range 213--369
K. In two different shock tube studies,\cite{mic92:3228,mic93:6389}
the rate was measured at 1850--3160 K and 1251--3152 K to be
$3.32\times10^{-11}$ and $3.7\times 10^{-11}$
cm$^3$s$^{-1}$molecule$^{-1}$, respectively. The low temperature rates
for this reaction is measured as $(3.2\pm0.6)\times 10^{-11}$
exp$(25\pm16/T)$cm$^3$s$^{-1}$molecule$^{-1}$ in a continuous
supersonic flow reactor at 48--211 K.\cite{ber09:8149} The recommended
value, suggested by Baulch et. al., for combustion modelling is
$3.5\times 10^{-11}$ cm$^3$s$^{-1}$molecule$^{-1}$ over the
temperature range 210--3700 K.\cite{bau05} Most of the experiments
mentioned above suggest high rates for the forward reaction at low
temperatures which points towards a barrierless process.\\

\noindent
For the reverse reaction only few experimental studies have been
reported. The rate expression obtained from a shock tube experiment at
2384--3850 K by heating N$_2$/O$_2$/N$_2$O/Kr mixtures\cite{mon79:543}
was given as $1.84\times10^{14}$exp$(-76250/RT)$
cm$^3$s$^{-1}$mol$^{-1}$ (corresponding to
$3.055\times10^{-10}$exp$(38370/T)$
cm$^3$s$^{-1}$molecule$^{-1}$). Another, later,
experiment\cite{thi85:685} determined the rate coefficient from O- and
N-concentration measurements in shock heated N$_2$/N$_2$O/Ar mixtures
at 2400--4100 K to be $3.0\times 10^{-10} \exp{(-38300/T)} \pm 40 \%$
cm$^3$s$^{-1}$molecule$^{-1}$. The rate of NO formation in the burned
gas region of high temperature oxypropane flame has been estimated
from experiment and using a mathematical modelling at 2880 K as
$10^{7.93}$ cm$^3$s$^{-1}$mol$^{-1}$ (corresponding to
$1.413\times10^{-16}$
cm$^3$s$^{-1}$molecule$^{-1}$).\cite{Livesey:1971} \\

\noindent
Theoretical studies focused primarily on exploring the electronic
structure for the system, constructing potential energy surfaces
(PESs) and carrying out dynamical simulations on these PESs. {\it Ab
  initio} PESs have been determined for the $^3$A$'$ and $^3$A$''$
states of N$_2$O using complete active space self consistent field
(CASSCF)/contracted CI (CCI) calculations.\cite{wal87:6946} For the
forward reaction, a small barrier of 0.5 kcal/mol was found on the
$^3$A$''$ PES whereas a considerably larger barrier of 14.4 kcal/mol
was found on the $^3$A$'$ PES. For the reverse reaction the
endoergicity was calculated to be 75 kcal/mol. An analytical PES for
the $^3$A$''$ state of N$_2$O was constructed from CASSCF/CCI energies
using a Sorbie-Murrell functional form and quasiclassical trajectory
(QCT) calculations were performed for the N+NO collisions at 300
K.\cite{gil92:5542}\\

\noindent
Although the triplet states of N$_2$O have immense importance in
environment, atmospheric chemistry and high energy collisions in
hyperthermal flow, only three
PESs\cite{gam03:2545,lin16:024309,alp17:2392} are available in the
literature. More recent electronic structure calculations have been
performed on both the triplet states of N$_2$O using CASSCF, CASPT2
methods and different basis sets.\cite{gam03:10602} Later the
CASPT2/cc-pVTZ, energies from Ref. \citen{gam03:10602} were used along
with additional energies at the same level of theory to construct
analytical representations for the $^3$A$'$ and $^3$A$''$
PESs.\cite{gam03:2545} The root mean square deviations (RMSD) of these
fits were 2.28 and 1.8 kcal/mol for the $^3$A$'$ and $^3$A$''$ PESs,
respectively. This work confirms that there is no barrier to form
N$_2$+O when approaching from the N+NO side for the $^3$A$''$
state. The rates for the forward and the reverse reactions were
calculated from improved canonical variational transition-state (ICVT)
theory on the new PESs and and also on the PESs from
Ref. \citen{gil92:5542}.  Quantum wave packet dynamics had been
carried out to study the N+NO reaction on both triplet states and
rates were calculated for a wide range of temperatures using the
$J$-shifting approximation.\cite{gam03:7156,gam06:174303} The forward
reaction was studied using QCT to obtain the rates upto 5000 K and
explore the energy distributions and reaction
mechanisms.\cite{gam10:144304} Also, the reverse reaction was studied
again via QCT on the CASPT2/cc-pVTZ PESs\cite{gam03:2545} to obtain
the vibrational relaxation rates from $v=1$ to $v'=0$
states.\cite{esp17:6211}\\
 
\noindent
Analytical PESs were also constructed for the two lowest triplet
states of N$_2$O based on MRCI/maug-cc-pVTZ calculations and applying
the dynamically scaled external correlation method with
root-mean-squared errors (RMSE) of 4.44 and 3.71 kcal/mol for the
$^3$A$'$ and $^3$A$''$ PESs, respectively.\cite{lin16:024309}. Six
states were included in the dynamically weighted state-averaged CASSCF
calculations and permutational invariant polynomials were used to
represent the many-body part of the PESs.  Geometries and energetics
of the stationary states were found to differ from the earlier
PESs\cite{gam03:2545}. QCT calculations for the O+N$_2$ collisions at
high energies\cite{lin16:234314} on the triplet PESs of
Ref. \citen{lin16:024309}\\

\noindent
Finally, a reproducing kernel based PESs for the $^3$A$'$ and
$^3$A$''$ states were calculated using MRCI+Q/cc-pVTZ energies and
rates and product state distributions were calculated up to 20000
K.\cite{alp17:2392} However, due to presence of a small barrier
  on the $^3$A$''$ PES in the N+NO channel the calculated QCT rates
were smaller at low temperatures. Recently, a state-to-state cross
section model for the forward reaction based on the $^3$A$'$ PES has
been developed using neural network which predict the thermal rates
and product state distributions accurately.\cite{MM.nn:2019} \\

\noindent
Reaction rates obtained from the PESs using fits to the CASPT2/cc-pVTZ
data\cite{gam03:2545} are in good agreement with the
experiments.\cite{gam03:2545,gam06:174303,gam10:144304} However, the
vibrational relaxation rates for the reverse reaction are
significantly smaller than the experimental results up to 7000
K.\cite{esp17:6211} The PESs in Ref. \citen{gam03:2545} were
constructed using a lower level of theory (CASPT2/cc-pVTZ) and it is
found that number and geometry of the stationary states differs using
higher level (MRCI/maug-cc-pVTZ) of theory and more states in the
CASSCF calculations.\cite{lin16:024309} These PESs are meant for
exploring the dynamics of the O+N$_2$ collisions at high
energies. Since the forward reaction is barrierless and is highly
reactive at very low temperatures, a very reliable description of the
N+NO channel is required in the asymptotic region, which was absent in
Ref. \citen{lin16:024309} PESs. On the other hand the reproducing
kernel Hilbert space (RKHS) interpolation procedure can capture the
correct long range behavior for different types of
interactions.\cite{MM.rkhs:2017,MM.cno:2018,MM.heh2:2019,MM.no2:2020}\\

\noindent
Another relevant quantity that balances the energy content of hot gas
flow is the exchange and relaxation of vibrational energy which can
also be determined from QCT simulations.\cite{Nikitin:07,vrthesis}
This property has been found to be sensitive to different regions of
the PES than thermal rates\cite{MM.no2:2020} and provides an
additional possibility to validate the quality of the PESs. Rates for
vibrational relaxation serve also as input data for more coarse
grained simulations, such as direct simulation Monte Carlo
(DSMC)\cite{dsmc}. With N$_2$ and O$_2$ as two major components of
air, N$_2$+O inelastic collisions may significantly affect the energy
content and energy redistribution in high temperature air flow. The
vibrational relaxation (VR) rates are calculated from relaxation time
parameter ($p\tau_{\rm vib}$) using the Bethe-Teller
model.\cite{btmodel,vrthesis} \\

\noindent
Here, global PESs have been constructed for the $^3$A$'$ and $^3$A$''$
electronic states for the reactive [NNO] system using RKHS
interpolation from a large number ($> 10 000$) of MRCI+Q/aug-cc-pVTZ
energies for each state. The present RKHS PESs are constructed using
$\sim 3$ times more {\it ab initio} energies than the previous
PESs\cite{lin16:024309} and thus cover all the important interaction
regions. Using a RKHS allows one to impose the correct long range
behavior for the dipole-induced dipole interactions ($\sim 1/R^6$) and
also reproduce the topology of the {\it ab initio} PESs
accurately. Quasiclassical trajectories are then run on these new PESs
to determine thermal and vibrational relaxation rates which allow
validation of the PESs by comparing with experiments.\\

\noindent
The present article is organised as follows. The methodological
details of constructing the RKHS PESs and QCT dynamics are discussed
in Section II followed by presenting the results and discussing them
in Section III and finally concluding this work in Section IV.

\begin{figure}
\includegraphics[scale=1.3]{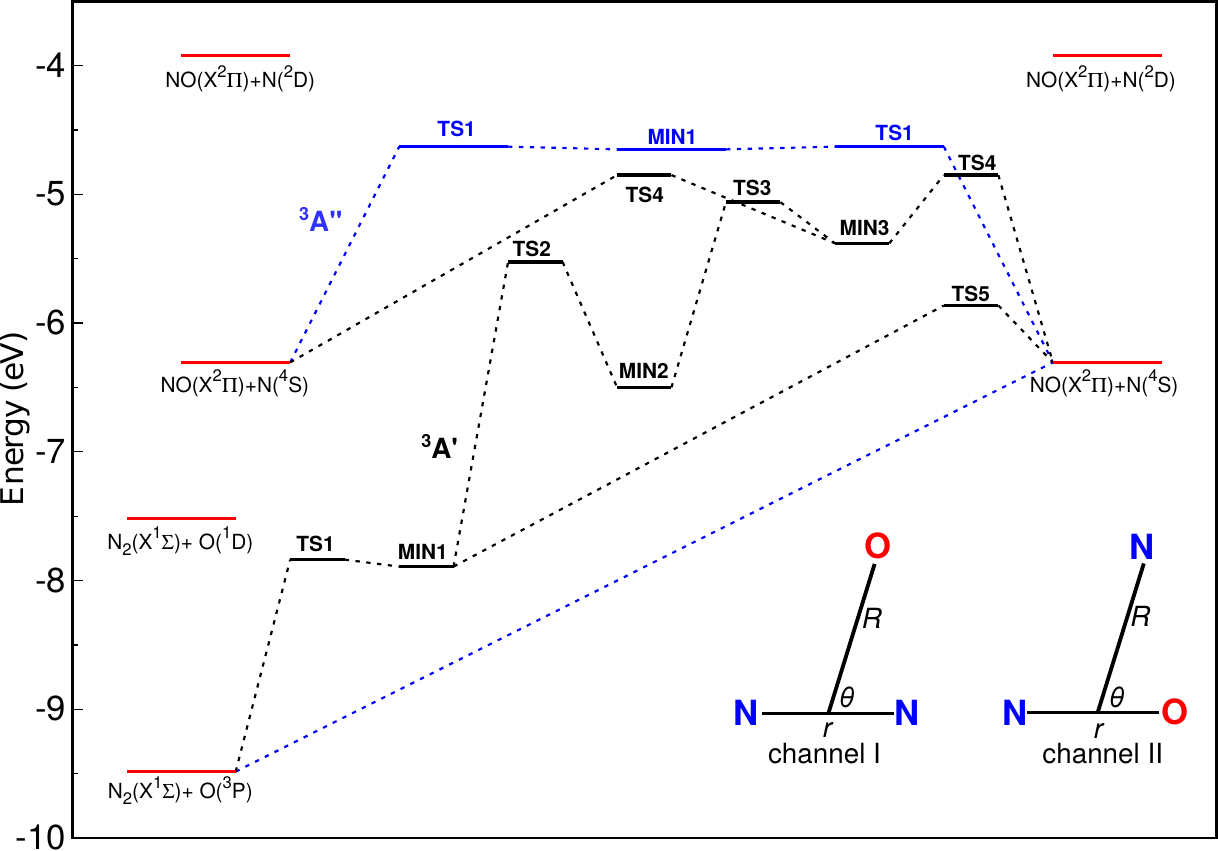}
\caption{Schematic energy diagram for the N$_2$O system showing
  different reactant channels, minima (MIN$i$) and transition states
  (TS$i$).  The black lines connect the configurations on the $^3$A$'$
  while the blue lines connect the stationary states on $^3$A$''$
  states. Two grids in Jacobi coordinates as defined for the two
  channels for the {\it ab initio} energies are shown at bottom
  right.}
\label{fig:fig1}
\end{figure}

\section{Methods}
\noindent
This section summarizes the methods for calculating the {\it ab
  initio} energies for the diatomic and triatomic systems,
construction of the analytical representation of the {\it ab initio}
PESs using RKHS, quasiclassical trajectory calculations on those PESs
and finally calculation of reaction and vibrational relaxation
rates.\\

\subsection{Electronic structure calculations}
\noindent
{\it Ab initio} energy calculations for the triatomic system were
performed on a grid defined in Jacobi coordinates $(r,R,\theta)$ where
$r$ is the diatomic separation, $R$ is the distance between the center
of mass of the diatom and the free atom and $\theta$ is the angle
between $\vec{r}$ and $\vec{R}$. For the present case, two separate
grids were used for the (I) O+N$_2$ and (II) N+NO channels,
respectively. The Jacobi coordinates for the two reactive systems are
shown in Figure \ref{fig:fig1}. For $R$, the grid included 30 and 28
points from 1.4 to 12.0 a$_0$ for the two channels. The diatomic
separation $r$ for channels I and II are covered by 20 points from
1.55 to 4.0 a$_0$ and 21 points, from 1.50 to 4.0 a$_0$,
respectively. For the angular grid Gauss-Legendre quadrature points
were used between $0^\circ$ to $180^\circ$. The quadrature points
covered the angular domain more efficiently than the regular grid as
in the RKHS scheme, the angular grid is transformed to a new
coordinate according to $\frac{1 - {\rm cos} \theta}{2}$ into a grid
from 0 to 1 which leads to a denser grid near 0 or 1 and evenly spaced
in between. In the present case, 13 quadrature points define the
Jacobi angle which are $169.796^\circ$, 156.577$^\circ$,
143.281$^\circ$, 129.967$^\circ$, 116.647$^\circ$, 103.324$^\circ$,
90.100$^\circ$, 76.676$^\circ$, 63.353$^\circ$, 50.033$^\circ$,
36.719$^\circ$, 23.423$^\circ$ and 10.204$^\circ$.\\

\noindent
The $C_s$ symmetry, which is the highest symmetry for the present
system to describe all geometries, was used to perform the electronic
structure calculations.  The mutlireference configuration interaction
with Davidson correction (MRCI+Q)\cite{davidson:1974,wer88:5803}
method and augmented Dunning-type correlation consistent polarized
triple zeta (aug-cc-pVTZ)\cite{dun89:1007} basis set were used to
calculate the electronic structure for all configurations considered
in this work. This level of theory has been found to describe the
electronic structures for the C-, N-, O-containing species quite well
for the entire region of the global
PES.\cite{lin16:024309,MM.cno:2018,MM.no2:2020} State-averaged
complete active space self-consistent field
(CASSCF)\cite{wen85:5053,kno85:259,wer80:2342,werner:2019}
calculations were carried out prior to the MRCI calculations in order
to obtain a smooth topology of the MRCI PESs. In the State-averaged
CASSCF calculations a total of eight states were included (the two
lowest states from each spin (singlet and triplet) and spatial (A$'$
and A$''$) symmetry).  Including all these eight states in the CASSCF
calculations provides consistency in the energies and numerically
stabilizes the convergence for configurations with closely-lying
states of the same symmetry. All the electronic structure calculations
are performed using the Molpro-2019.1 \cite{MOLPRO_brief} software
package.\\

\noindent
For each of the electronic states, 4200 and 7644 {\it ab initio}
electronic structure calculations have been carried out for the
N$_2$+O and N+NO channel, respectively. Here it is worth to mention
that $\sim 10$ \% of the electronic structure calculations converged
to the excited states in the long range interaction regions (i.e., for
large values of $R$ and/or $r$). These energies are excluded from the
grid to construct the RKHS. A small number ($< 0.5$ \%) of {\it ab
  initio} calculations did not converge at all. For both these cases
the missing grid points were computed using a 2-dimensional
reproducing kernel ($R,r$) for a particular $\theta$ before inserting
them into the final 3D RKHS. Thus $\sim 3$ times more reference {\it
  ab initio} energies have been used in this PESs than the previous
PESs for this system.\cite{lin16:024309}\\

\subsection{Reproducing Kernel Representation of the PES}
The reproducing kernel Hilbert space (RKHS) interpolation
technique\cite{ho96:2584} has been used to construct the analytical
representation of the PESs for both electronic states ($^3$A$'$ and
$^3$A$''$). In the RKHS approach, the approximated value
$\tilde{f}({\bf{x}})$ of a function $f({\bf x})$ is computed from a
set of known values $f({\bf x}_i)$ as a linear combination of kernel
polynomials, where ${\bf x}$ are variables e.g. set of spatial
coordinates. The RKHS procedure is documented in detail in
Refs. \citen{ho96:2584,MM.rkhs:2017} and has also been used to
construct PESs for other triatomic
systems.\cite{MM.cno:2018,MM.heh2:2019,MM.no2:2020}\\

\noindent
In order to well describe the atom+diatom dissociation channels the
total potential $V(R,r,\theta)$ energy of N$_2$O is expanded as
\begin{equation}
 V(R,r,\theta) = E(R,r,\theta) - V(r)
 \label{mbe}
\end{equation}
where $V(r)$ is the diatomic potential for the respective channel.
For this purpose, the {\it ab initio} energies for NO and N$_2$ are
calculated separately and represented as a 1D RKHS using reciprocal
power decay kernel $k^{2,6}(r,r')$. The values of $E(R,r,\theta)$ are
then calculated from Eq. \ref{mbe} and $E(R,r,\theta) \rightarrow 0$
for $R \rightarrow \infty$. As the values of the radial kernel
function approach zero at large distances, $E(R,r,\theta)$ can be
represented by a 3D kernel as
\begin{equation}
     K({\bf{x}}, {\bf{x}}') =  k^{2,6}(R,R') k^{2,6}(r,r')  k^{2}(z,z').
\end{equation}
Reciprocal power decay kernels are used for the radial dimensions ($R$
and $r$). For large separations they approach zero according to
$\propto \frac{1}{x^n}$ (here $n=6$) which gives the correct
long-range behavior for neutral atom-diatom type interactions. For the
angular coordinate a Taylor spline kernel $k^{2}(z,z')$ is used where
$z = \frac{1 - {\rm cos} \theta}{2}$. The definitions of the kernel
functions can be found in Ref. \citen{ho96:2584,MM.rkhs:2017}.  The
RKHS procedure exactly reproduces the values of the function at the
known points for smooth functions. However, regularization parameter
(a very small value, $10^{-19}$ is used in this work) is often used to
reduce the noise in the data set and to numerically stabilize the
algorithm.\\

\noindent
The global PES is constructed by mixing three PESs for each channels
(note there are two channels for N+NO) multiplying a smooth weight
function $w_i({\bf r})$ which is expressed as
\begin{equation}
 w_i({\bf r})=\frac{e^{-(r_i/dr_i)^2}}{\sum_{j=1}^{3}e^{-(r_j/dr_j)^2}},
\end{equation}
where ${\bf r}$ are the three internuclear distances ($r_i$ represents
each diatomic bond length) and $dr_j$ are parameters for each
channels which are optimized by least square fitting.\\

\subsection{Quasiclassical trajectory calculation}
The N+NO and O+N$_2$ collisions systems are studied by running
quasiclassical trajectories (QCT) on the global RKHS PESs for both
$^3$A$'$ and $^3$A$''$ states of N$_2$O. The QCT method followed here
is well documented in Refs. \citen{kar65:3259,tru79,hen11} and
\citen{konthesis}.  In this approach, Hamilton's coupled differential
equations of motion are solved using the fourth-order Runge-Kutta
method.  The integration time step was $\Delta t = 0.05$ fs which
guaranteed conservation of the total energy and angular momentum up to
the sixth and eighth decimal places, respectively. Initial conditions
for the trajectories are sampled using a Monte Carlo sampling
method. The ro-vibrational states for the reactant and product are
computed using the semiclassical theory of bound states. The
ro-vibrational states for the product diatom are assigned to the
nearest integer using histogram binning (HB) and Gaussian binning (GB)
schemes.\cite{bon97:183,bon04:106,konthesis} Since the results from
both binning schemes are similar, only those obtained from HB are
discussed unless otherwise mentioned.\\

\noindent
The probability of an event $x$ (reactive, vibrational relaxation) can
be computed as
\begin{equation}
P_x = \frac{N_x}{N_{\rm tot}},
\end{equation}
where $N_x$ is the number of trajectories corresponding to the event
of interest and $N_{\rm tot}$ is the total number of trajectories. The
cross section for that event is then computed as
\begin{equation}
\sigma_x = \pi b^2_{\rm max}P_x
\end{equation}
where $b_{\rm max}$ is the maximum impact parameter for which the
event can occur. The impact parameters are sampled using stratified
sampling\cite{tru79, ben15:054304,MM.cno:2018} by subdividing the
interval between 0 and $b_{\rm max}$ into six strata.\\

\noindent
The rates for an event $x$ at a particular temperature is then
obtained as
\begin{equation}
 k_{i,x}(T) = g_i(T)\sqrt{\frac{8k_{\rm B}T}{\pi\mu}} \pi b^2_{\rm max}P_x.
\end{equation}
where $g_i(T)$ is the electronic degeneracy factor of an electronic
state `$i$', $k_{\rm B}$ is the Boltzmann constant and $\mu$ is the
reduced mass of the collision system.\\

\noindent
For the N($^4$S) + NO(X$^2\Pi$) collision process, the temperature
dependent degeneracies for both triplet states are
\begin{equation}
g^f(T)=\frac{3}{4\cdot(2+2\cdot e^{\frac{-177.1}{T}})}
\label{ge1}
\end{equation}
whereas for the reverse O($^3P$) + N$_2$($X^1\Sigma$) collision
process the degeneracies are
\begin{equation}
g^r(T)=\frac{3}{1\cdot(5+3 \cdot e^{\frac{-227.8}{T}}+1\cdot
  e^{\frac{-326.6}{T}})}
  \label{ge2}
\end{equation}
The numbers in Eqs. \ref{ge1} and \ref{ge2} are the degeneracies of
the $J$ or spin states while the exponential parameters 177.1 K, 227.8
K and 326.6 K are the energy gaps between two $J$ states. The
equilibrium constant is calculated as
\begin{equation}
K_{eq}(T)=\frac{k_{f}(T)}{k_{r}(T)},
\label{equ}
\end{equation}
where, $k_{f}(T)$ and $k_{r}(T)$ are the rates for the forward and the
reverse reaction.  In the rate calculations, for each trajectory the
initial relative translational energies of the reactants are sampled
from a Maxwell-Boltzmann distribution and ro-vibrational states of the
reactant diatom is sampled from a Boltzmann distribution at
temperature $T$.\cite{tru79,MM.cno:2018}\\

\section{Results and Discussion}
\noindent
In the following, the topology and accuracy of the PESs, the thermal
rates and the rates for vibrational relaxation from QCT simulations
are presented and discussed.

\subsection{Topology and Accuracy of the Potential Energy Surfaces}
\begin{figure}[h!]
\includegraphics[scale=0.7]{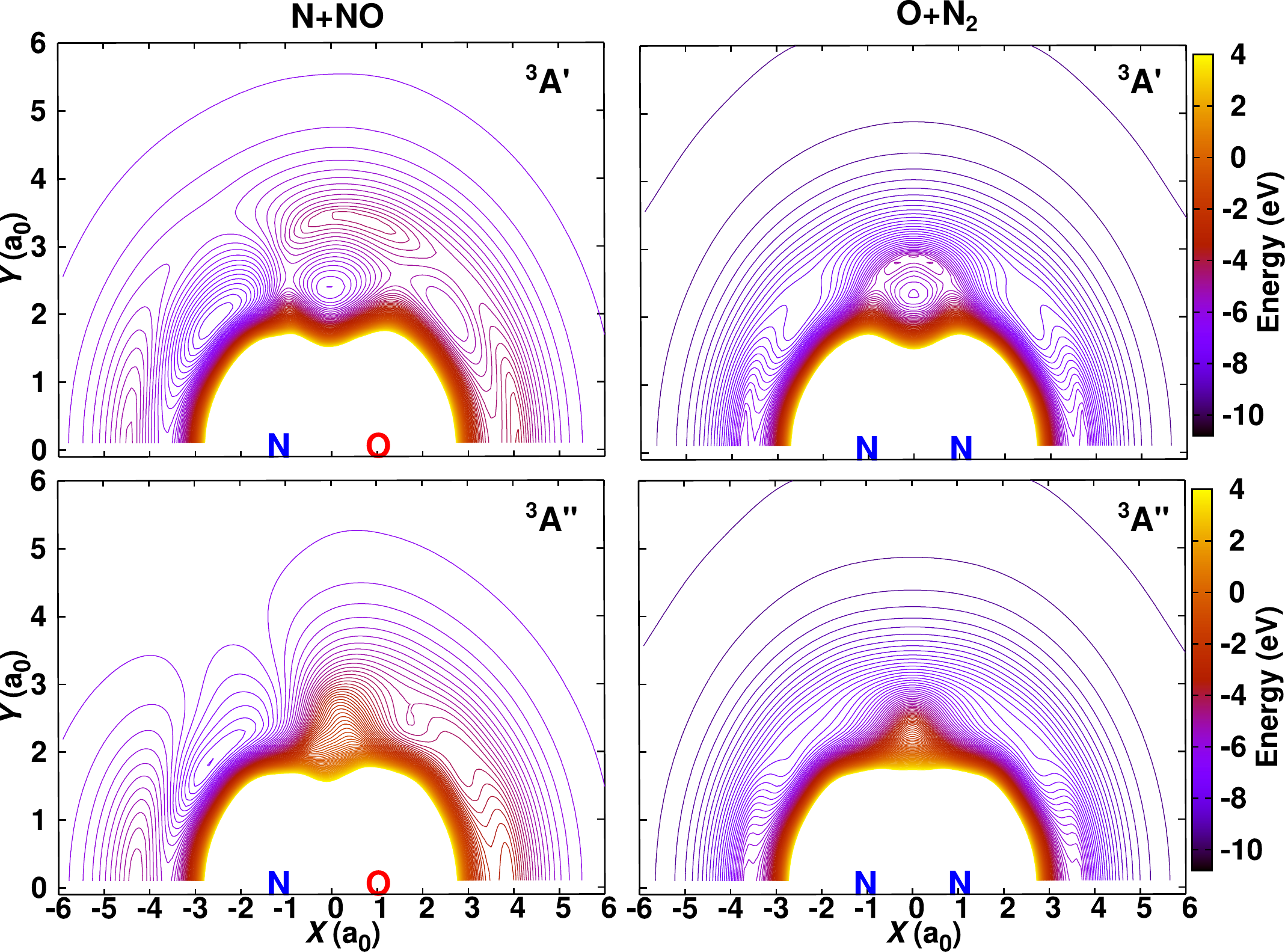}
\caption{Contour plots of the RKHS interpolated energies shown in
  relaxed PES representation (see text) for different electronic
  states. The diatoms are on the `X' axis while the origin of each
  plot is set to the center-of-mass of the diatoms. Spacing between the
  contour lines is 0.15 eV. Zero of energy is set to the atomization
  energy of N$_2$O i.e., the energy of N($^4$S)+N($^4$S)+O($^3$P).}
\label{fig:fig2}
\end{figure}

\noindent
The topography of the $^3$A$'$ and $^3$A$''$ PESs for the two
collision systems N+NO and O+N$_2$ - is reported in Figure
\ref{fig:fig2} which shows $r$-relaxed PESs. The center of mass of the
diatom is placed at the origin and the position of the third free atom
is described as $(x,y)$ on a 2D space. These 2D PESs are computed by
determining the minimum energy for a given $(x,y)$ with $r \in [2.0,
  2.5]$ a$_0$ for N+NO and $r \in [1.9, 2.4]$ a$_0$ for O+N$_2$. This
was done because a `relaxed' PES better represents all possible stationary
states than a conventional 2D PES with a fixed $r$
value.\cite{var87:455}\\

\noindent
Both PESs have significantly different anisotropic topologies
depending on the angle at which the reactants approach one
another. The topology of the $^3$A$'$ PES is highly structured
compared with that of the $^3$A$''$ PES for both channels. For both
PESs, the global minimum is at the O+N$_2$ asymptote while the N+NO
asymptote is higher in energy by 73.1 kcal/mol. For the N+NO
collisions, reactants face barriers for all angular approach for the
$^3$A$'$ PES. The barrier height $\sim 9.6$ kcal/mol (equivalent to
0.4 eV or $\sim 4800$ K) is lowest for $\sim 135^\circ$ approach. On
the other hand, the N+NO collisions are barrierless on the $^3$A$''$
PES for $\sim 135^\circ$ which will profoundly affect the reactivity
at lower temperatures. A potential well can be seen on the $^3$A$'$
PES along the near perpendicular approach (MIN2, see Figure
\ref{fig:fig1}) which is not present on the $^3$A$''$ PES. For both
PESs, the O+N$_2$ collisions have highly repulsive landscapes and a
deep potential well with a $C_{2v}$ symmetry at high energies on the
$^3$A$'$ PES along the perpendicular approach (MIN2, see Figure
\ref{fig:fig1}).\\

\noindent
The stationary states on the PESs are characterized and presented in
Figure \ref{fig:fig1} and tabulated in Table \ref{tab:mints}. The
minima and transition states are calculated by using the BFGS and
nudged elastic band methods\cite{hen00:9901,kol16:094107} implemented
in the Atomic Simulation Environment (ASE) package.\cite{ase} The N+NO
$\rightarrow$ N$_2$+O reaction is barrierless on the $^3$A$''$ PES
while the N exchange N$_{\rm A}$+N$_{\rm B}$O $\rightarrow$ N$_{\rm
  A}$O+N$_{\rm B}$ reaction has a barrier of 38.03 kcal/mol which is
$\sim 4.6$ and $\sim 2.4$ kcal/mol smaller than for the earlier
PESs\cite{lin16:024309,gam03:2545}, respectively. A very shallow well
($\sim 0.5$ kcal/mol) with a $C_{2v}$ symmetry exists as MIN1 between
two TS1 structures.\\

\noindent
There are two pathways from N+NO to N$_2$+O on the $^3$A$'$ PES which
are similar to Ref. \citen{lin16:024309}. The minimum energy path
follows N+NO $\rightarrow$ TS5 $\rightarrow$ MIN1 $\rightarrow$ TS1
$\rightarrow$ N$_2$+O. TS5 is located 9.58 kcal/mol higher than the
N+NO asymptote which compares with 10.50 kcal/mol and 8.35 kcal/mol
for the PESs in Ref. \citen{lin16:024309} and \citen{gam03:2545},
respectively.  However, MIN1 and TS1 are not present on the $^3$A$'$
PES in Ref. \citen{gam03:2545}. MIN1 is located at 35.86 kcal/mol
higher than the O+N$_2$ asymptote and has a depth of 1.25 kcal/mol
from TS1. The NN bond is shorter than the NO bonds in MIN1. The N
exchange path for the N$_{\rm A}$+N$_{\rm B}$O $\rightarrow$ N$_{\rm
  A}$O + N$_{\rm B}$ reaction has a barrier of 32.93 kcal/mol (TS4)
and a potential well with a $C_{2v}$ symmetry of 12.23 kcal/mol. TS3
has $C_{2v}$ symmetry and it connects MIN3 and MIN2 ($C_{2v}$ geometry
and close to an equilateral triangle with angles 61.8$^\circ$,
59.1$^\circ$ and 59.1$^\circ$). MIN2 is connected with MIN1 via TS2.\\

\begin{figure}[h!]
\includegraphics[scale=1.8]{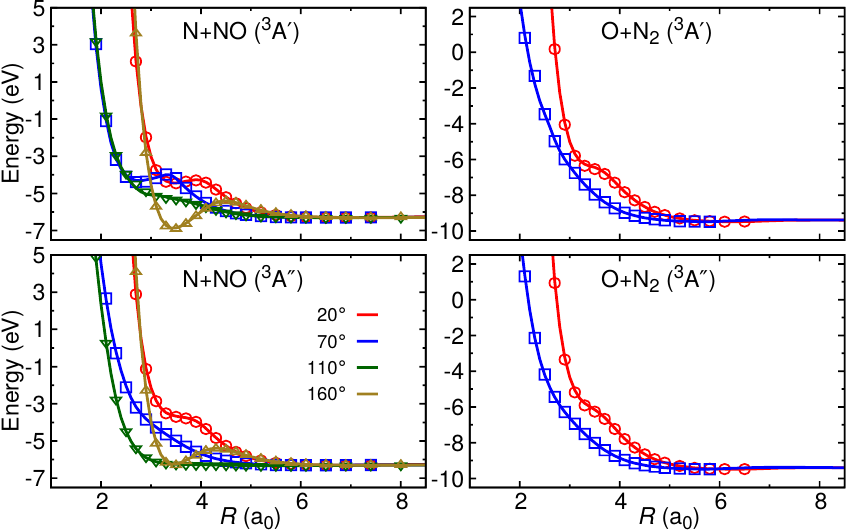}
\caption{Comparison between the {\it ab initio} energies (open
  symbols) (not part of the training grid) and the RKHS interpolated
  energies (solid lines) in Jacobi coordinates along the N+NO and
  O+N$_2$ channels for different angles.  The diatomic bond distances
  for NO and N$_2$ are fixed at 2.19 and 2.08 a$_0$,
  respectively. Zero of energy is set to the atomization energy of
  N$_2$O i.e., energy of N($^4$S)+N($^4$S)+O($^3$P).}
\label{fig:fig3}
\end{figure}

\noindent
To verify the quality of the RKHS PESs, the correlation between the
{\it ab initio} and the RKHS energies for both states for the training
grid are reported in Figure S1 upper panel. The
structures with energy 6.8 eV above the N+N+O asymptote 7908 and 7797
points for the training grid) have $R^2$ values of 0.99995 and 0.99996
and RMSE of 0.509 and 0.456 kcal/mol for the $^3$A$'$ and $^3$A$''$
states, respectively. Electronic structure calculations have also been
performed for off-grid, randomly generated geometries for both states
which serve as an additional test data set. A correlation diagram for
the test grid is shown for 494 and 497 points for the $^3$A$'$ and
$^3$A$''$ states, respectively, with energies $\leq 6.8$ eV (see
Figure S1 lower panel). The correlation coefficients
obtained for the test grids are 0.99985 and 0.99988 with RMSE 0.82 and
0.80 kcal/mol for the $^3$A$'$ and $^3$A$''$ states, slightly larger
than for the reference points but still of high quality. The RMSE
values for the present PESs are more than three times lower than for
previous PES.\cite{gam03:2545,lin16:024309}\\

\noindent
The quality of the RKHS PESs developed in this work is further
examined by considering 1D cuts along $R$ for different values of
$\theta$ and constant values of $r$ for the N+NO and O+N$_2$ channels,
see Figure \ref{fig:fig3}. Again, these are off-grid cuts (not part of
the training grid) and the
agreement between the {\it ab initio} and RKHS interpolated energies
is excellent for all the cuts.\\

\begin{table}[h!]
  \caption{Geometries and energies for the minima (MIN$i$) and
    transition states (TS$i$) on the $^3$A$'$ and $^3$A$''$ PESs of
    N$_2$O. Distances given in a$_0$, angles in degree and energies in
    eV for $\Delta E_1$ and in kcal/mol for $\Delta E_2$ (in kcal/mol)
    with respect to the energy of N($^4$S)+N($^4$S)+O($^3$P). For the
    energy level diagram and the connectivity, see Figure
    \ref{fig:fig1}.}
	\begin{center}
	\begin{tabular}{l|rrr|r|r}
	\hline
	\hline
     & $R_{\rm NN}$ & $R_{\rm NO}$ & $\angle$NON & $\Delta E_1$  & $\Delta E_2$ \\
	\hline
		$^{3}$A$'$&\multicolumn{5}{c}{}\\
		\hline
	TS1   & 2.21 & 2.72 & 27.1   &  --7.837 & --180.736   \\
	MIN1  & 2.30 & 2.41 & 26.7   &  --7.892 & --181.990  \\
	TS2   & 2.58 & 2.56 & 50.0   &  --5.525 & --127.419  \\
	MIN2  & 2.70 & 2.63 & 61.8   &  --6.499 & --149.880  \\
	TS3   & 3.86 & 2.56 & 97.5   &  --5.061 & --116.703  \\
	MIN3  & 4.38 & 2.46 & 125.9  &  --5.381 & --124.080  \\
	TS4   & 4.73 & 2.25 & 121.1  &  --4.850 & --111.848  \\
	TS5   & 3.72 & 2.19 &  47.5  &  --5.863 & --135.193  \\
	\hline
	$^{3}$A$''$&\multicolumn{5}{c}{}\\
	\hline
	MIN1 & 3.92 & 2.51 & 102.9 & --4.651  & --107.245  \\
	TS1  & 3.97 & 2.36 & 102.0 & --4.629  & --106.746  \\
	\hline
	N+NO & & 2.19 & & --6.278& --144.774\\
	O+N$_2$& 2.09 & & &--9.447&--217.850\\
	\hline
	\hline
  \end{tabular}
	\end{center}
\label{tab:mints}
\end{table}

\subsection{Reaction and Dissociation Rates}

\begin{figure}[h!]
\centering \includegraphics[scale=0.5]{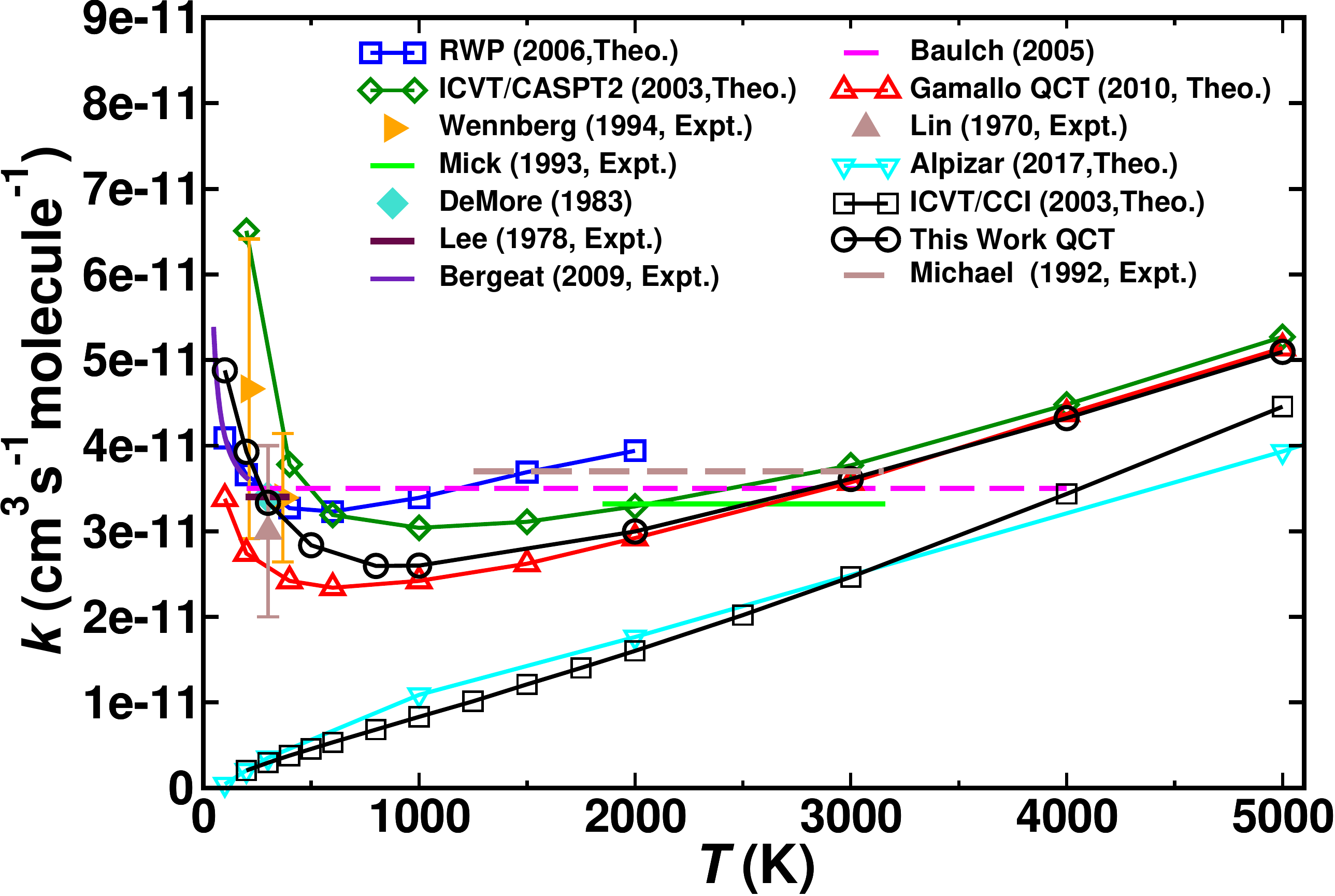}
\caption{Rate coefficients for the N($^4$S) + NO(X$^2\Pi$)
  $\rightarrow$ O($^3$P) + N$_2$(X$^1\Sigma$) from 100 to 5000 K.  The
  results calculated in the present work are shown as black open
  circles connected by solid black line.  Experimental (assigned as
  `Expt.') and Theoretical rates (assigned as `Theo.') available in
  the literature are
  shown.\cite{lee78:3069,dem83,mic93:6389,mic92:3228,wen94:18839,gam03:2545,gam06:174303,gam10:144304,alp17:2392,ber09:8149}}
\label{fig:fig5}
\end{figure}

\noindent
Total rate coefficients for the N($^4$S) + NO(X$^2\Pi$) $\rightarrow$
O($^3$P) + N$_2$(X$^1\Sigma$) reaction and the N exchange reaction
N$_{\rm A}$($^4$S) + N$_{\rm B}$O(X$^2\Pi$) $\rightarrow$ N$_{\rm
  B}$($^4$S) + N$_{\rm A}$O(X$^2\Pi$) have been calculated from 100 K
to 20000 K. These rates are based on 50000 quasiclassical trajectories
at each temperature. Individual contributions from each states are
tabulated in Tables S1 and S2,
respectively. Rate coefficients for N$_2$ formation (forward reaction)
are plotted between 100 and 5000 K in Figure \ref{fig:fig5}. In
addition to the present computations, those from previous experiments
and computations are shown. The rate is high at low temperature (100
K) and gradually decreases with increasing temperature before starting
to increase again though the variation is small over the entire
range. For the present work, the minimum and maximum rates are $2.595
\times10^{-11}$ and $5.095 \times10^{-11}$
cm$^3$s$^{-1}$molecule$^{-1}$ between 100 K and 5000 K, which differs
only by a factor of two, consistent with a barrierless reaction in the
forward direction.\\

\noindent
As can be seen in Table S1, due to presence of a
barrier of 9.58 kcal/mol in the entrance channel on the $^3$A$'$ PES
the contribution is considerably lower at low temperatures than that
of the $^3$A$''$ state which is barrierless. Hence, at low
temperatures the contribution from the $^3$A$''$ PES dominates.  Rates
for the forward reaction obtained from Ref. \citen{gam03:2545} using
PESs from Ref. \citen{gil92:5542,gil93:1719} (black open squares) and
from Ref. \citen{alp17:2392} (cyan open triangle down) are very small
at low temperatures and increase monotonically with increasing
temperature due to the presence of a small barrier on the $^3$A$''$
PES. In the present work, the results from QCT simulations correctly
describe the trends of the experimental results at low
temperature. However, the high temperature rate expression obtained
from experimental data are temperature independent whereas the QCT
rates increases slowly with temperature.\\

\noindent
Rates for the the N exchange reaction N$_{\rm A}$($^4S$) + N$_{\rm
  B}$O($X^2\Pi$) $\rightarrow$ N$_{\rm B}$($^4S$) + N$_{\rm
  A}$O($X^2\Pi$) are given in Table S2. Due to presence
of barriers along the reaction path on both PESs (see Figure
\ref{fig:fig1}), this channel only opens at $\sim 3000$ K (consistent
with a barrier height of $\sim 4800$ K, see above) and reactivity
increases with increasing temperature. However, the rates for the N
exchange reaction are smaller over the entire temperature range
considered.\\

\begin{figure}[h!]
\centering \includegraphics[scale=0.5]{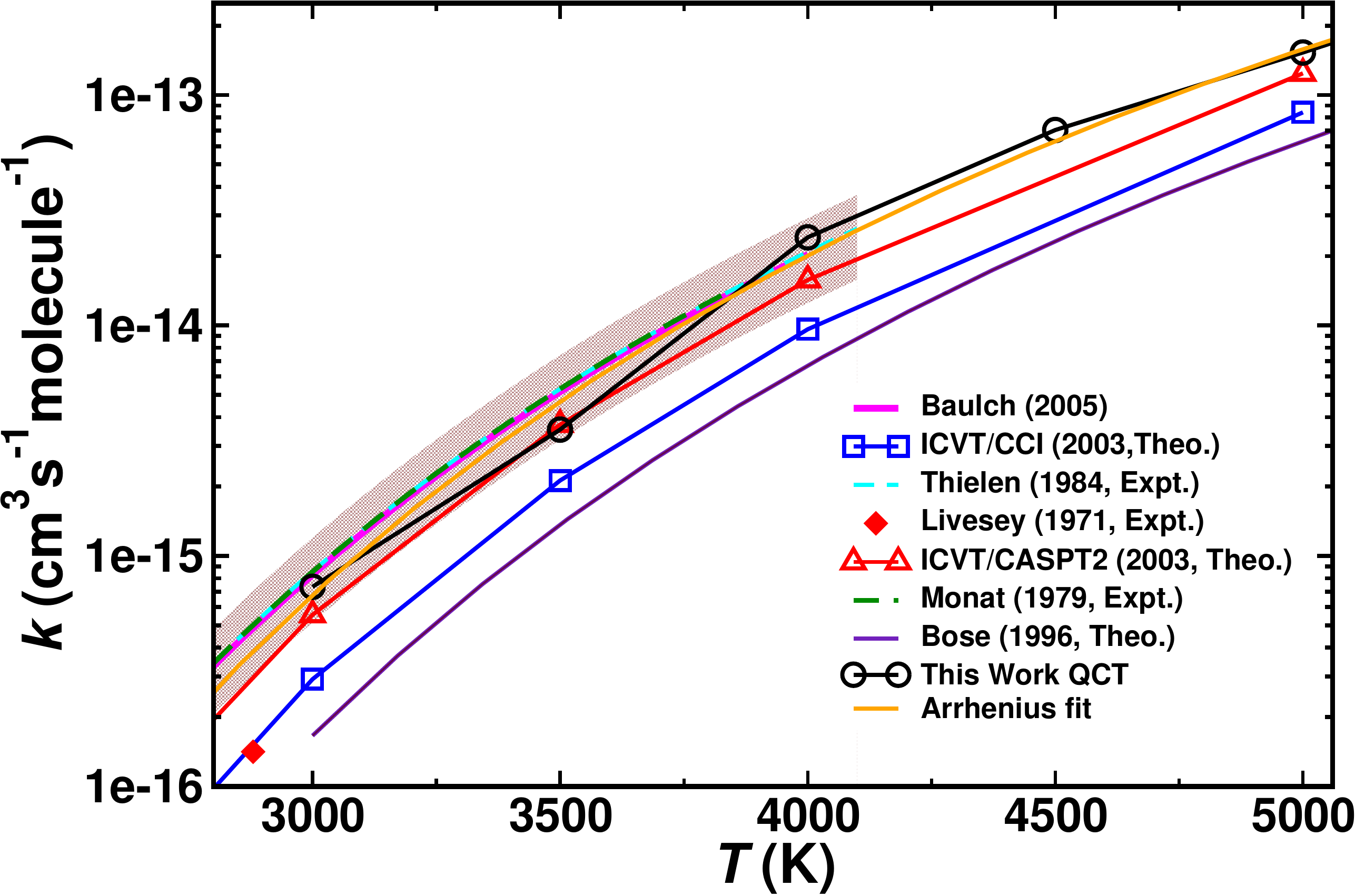}
\caption{Rate coefficients for the O($^3$P) + N$_2$(X$^1\Sigma$)
  $\rightarrow$ N($^4$S) + NO(X$^2\Pi$) reaction as a function of
  temperature. The results calculated in the present work are shown as
  black open circles connected by black line. Experimental and
  theoretical rates available from the literature are also
  shown.\cite{gam03:2545,bos96:2825,thi85:685,bau05,mon79:543,Livesey:1971} The
  shaded area shows the confidence limit for Thielen et. al. data.}
\label{fig:fig6}
\end{figure}

\noindent
Rate coefficients for the O($^3$P) + N$_2$(X$^1\Sigma$) $\rightarrow$
N($^4$S) + NO(X$^2\Pi$) reaction have been calculated from 2800 K to
20000 K by running $5\times10^4$ -- $4\times10^5$ quasiclassical
trajectories for each temperature. The rates are shown in Figure
\ref{fig:fig6} and individual contributions for each PES are given in
Table S3. This reaction is endothermic and starts at
high temperature $\sim 3000$ K. The reaction rate increases with
increasing temperature.  The contribution from $^3$A$'$ PES is less
than that from $^3$A$''$ PES in the entire range of temperature. Rates
reported from previous experimental and theoretical work are also
presented in Figure \ref{fig:fig6}. Rate expression obtained by
fitting experimental data are given in Ref. \citen{thi85:685} and
\citen{mon79:543} which are shown in Figure \ref{fig:fig6} along with
the recommended values provided by Baulch et al.\cite{bau05}. Results
obtained in the present work agree well with the experimental data.\\

\begin{figure}[h!]
\centering \includegraphics[scale=0.5]{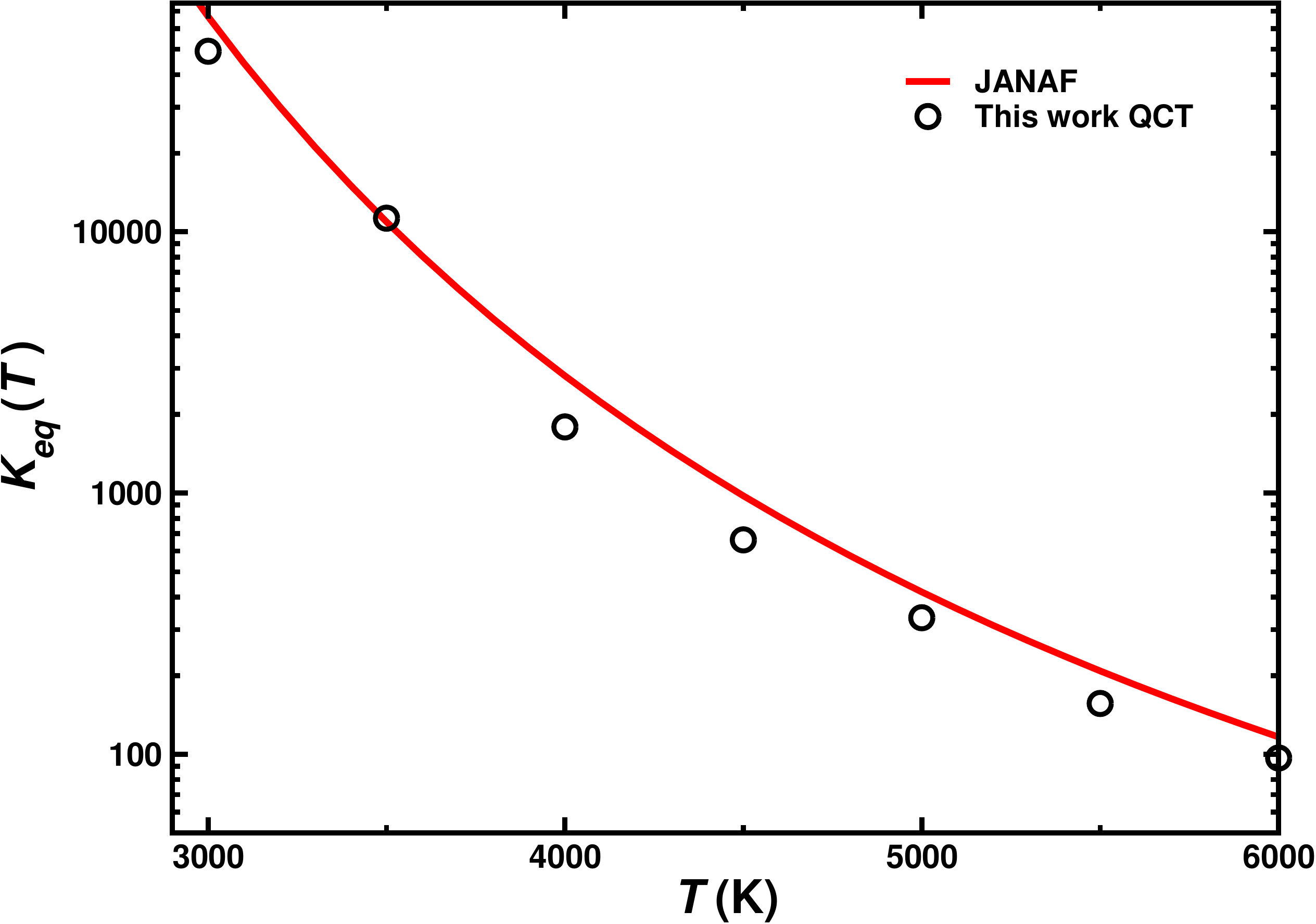}
\caption{Equilibrium constant $K_{eq}(T)$ for the N($^4S$) +
  NO($X^2\Pi$) $\leftrightarrow$ O($^3P$) + N$_2$($X^1\Sigma$)
  process.  Open circles represent the QCT results and the red line
  shows the results obtained from JANAF tables \cite{janaf}}
\label{fig:fig7}
\end{figure}

\noindent
The equilibrium constants for the N($^4$S) + NO(X$^2\Pi$)
$\leftrightarrow$ O($^3$P) + N$_2$(X$^1\Sigma$) reaction have been
computed using Eq. \ref{equ} between 3000 and 6000 K and compared with
the values reported in the JANAF\cite{janaf} tables. The QCT results
from this work are in good a agreement with the JANAF data computed from
thermodynamic quantities. This suggests that the present PESs are
accurate enough to describe the forward and the reverse reaction and
other electronic states play only a minor role in the dynamics of both
reactions.\\

\begin{figure}[h!]
\centering \includegraphics[scale=1.3]{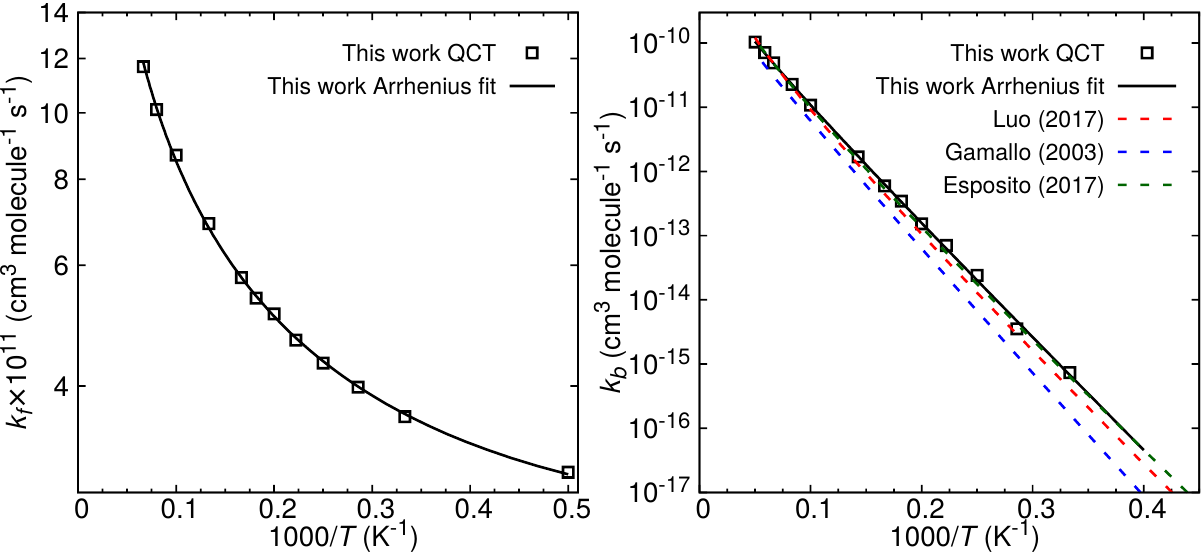}
\caption{Thermal rate coefficients for the N($^4$S) + NO(X$^2\Pi$)
  $\rightarrow$ O($^3$P) + N$_2$(X$^1\Sigma$) (left panel) and the
  reverse reaction (right panel) at temperatures relevant to
  hypersonic flight regime. Black open squares represent the QCT rates
  and the lines represent the Arrhenius fit. Rates calculated from the
  rate expression reported in literature are also shown as dashed
  lines.\cite{esp17:6211,gam03:2545,luo17:074303}}
\label{fig:fig5b}
\end{figure}

\noindent
Both N($^4S$) + NO($X^2\Pi$) $\rightarrow$ O($^3P$) +
N$_2$($X^1\Sigma$) and the reverse reaction are among the reactions
which play an important role at hyperthermal conditions during the
reentry of space vehicles into earth atmosphere. Analytical expression
for the rates are useful to simulate hypersonic flow during
reentry. In this work, the rates are calculated at hyperthermal
temperature for both reactions and modified Arrhenius functions
($k(T)=AT^{n}$exp$(-E_a/T)$) are fitted to those rates. The parameters
($A,n,E_a$) are given in Table \ref{tab:afit} and the fits are also
shown in Figure \ref{fig:fig5b} along with available rate expressions
from literature. For the reverse reaction, except for the fit from
Ref. \citen{gam03:2545} all the fits agree well over the entire
range.\\

\begin{table}[h]
\caption{Parameters obtained by fitting the rates to a modified
  Arrhenius equation. Rates in the 2000--15000 K range are used for
  the forward reaction while the rates from 3000--20000 K are used for
  the reverse reaction. Rate coefficients computed using these
  parameters have units in cm$^3$ molecule$^{-1}$ s$^{-1}$ with [$A$]
  = cm$^3$ molecule$^{-1}$ s$^{-1}$ and [$E_a$] = K while $n$ is
  unitless.}
\begin{tabular}{l|ccc}
\hline
\hline
Reaction & $A$ & $n$ & $E_a$  \\
\hline
Forward & $2.17214 \times 10^{-14}$ & 0.88796 & -946.02910 \\
Reverse & $7.73865 \times 10^{-12}$ & 0.46177 & 
39123.1149  \\
Dissociation & 4.55027 & -2.00227 & 129692.62993 \\
\hline
\hline
\end{tabular}
\label{tab:afit}
\end{table}

\begin{figure}[h!]
\centering \includegraphics[scale=1.2]{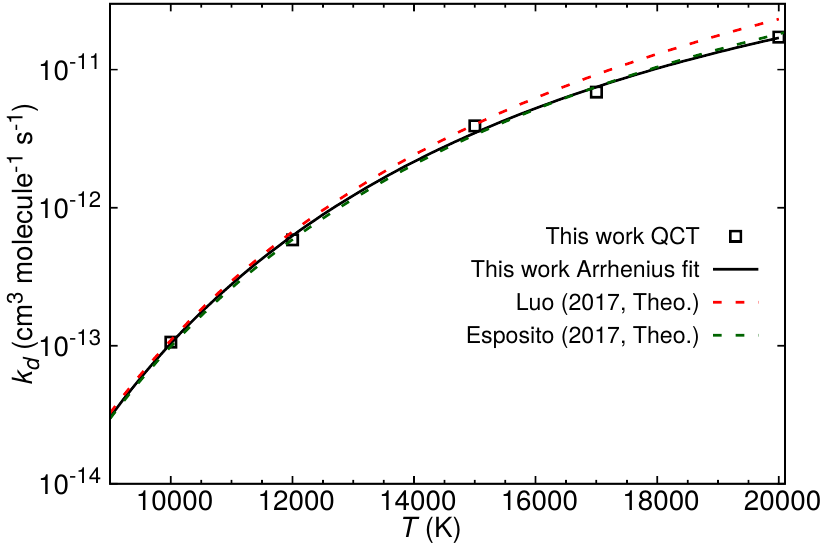}
\caption{Thermal rate coefficients for N$_2$ dissociation reaction for
  O($^3$P)+N$_2$(X$^1 \Sigma$) collisions. Black open squares represent
  the QCT rates and the line represent the Arrhenius fit. Rates
  calculated from the rate expression reported in
  Refs.\cite{luo17:074303,esp17:6211} are also shown as dashed lines.}
\label{fig:fig6a}
\end{figure}

\noindent
At very high temperatures collisions may lead to dissociation of the
diatomic molecules. As the dissociation energy for molecular oxygen is
considerably smaller than that of N$_2$, O$_2$ dissociates at
relatively low temperatures and generates atomic oxygen. At
hyperthermal conditions, dissociation of N$_2$ via N$_2$+O collisions
plays an important role to maintain the concentration of N
atoms. Here, the dissociation process of molecular nitrogen during
collisions with an oxygen atom is studied at high temperatures between
8000 K and 20000K. The rate coefficients for the N$_2$+O
$\rightarrow$2N+O are shown in Figure \ref{fig:fig6a} and also given
in Table S4. Fit to modified Arrhenius relation is also
plotted in Figure \ref{fig:fig6a} along with the rate expression
reported in Refs. \citen{luo17:074303} and \citen{esp17:6211}. The
parameters for the present fit are given in Table \ref{tab:afit}. All
the fits agree well except at high temperatures the fit from
Ref. \citen{luo17:074303} gives higher dissociation rate.\\

\subsection{Vibrational Relaxation}
\begin{figure}[h!]
\centering
\includegraphics[scale=1.5]{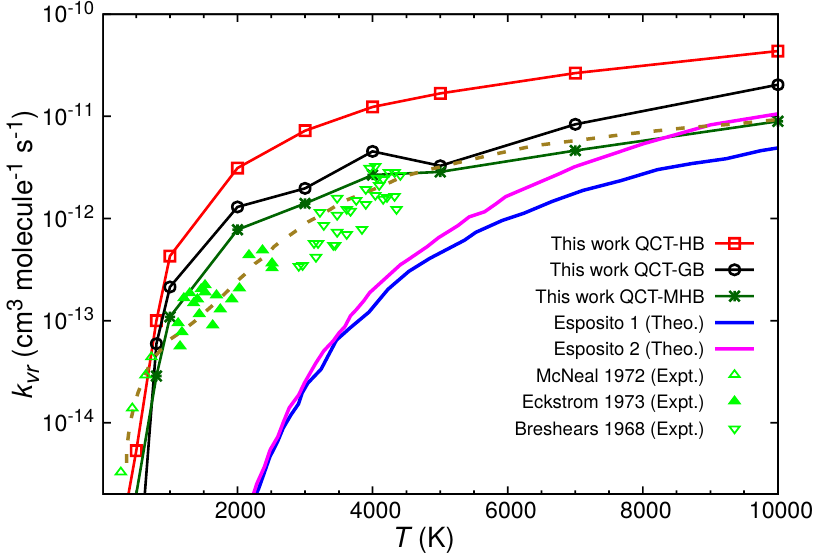}
\caption{Vibrational relaxation rates for O + N$_2$($\nu = 1$)
  $\rightarrow$ O+N$_2$($\nu' = 0$). Green symbols represent the
  experimentally determined VR
  rates.\cite{bre68:4768,mcn72:507,eck73:2787} Olive dashed line is
  the fit to the experimental result.\cite{on2vrbook} Rates obtained
  in this work from QCT simulations and using HB, GB and a modified HB
  scheme are given along with the full QCT (magenta solid line) and
  quasi-reactive QCT (blue solid line) results from
  Ref. \cite{esp17:6211}.}
\label{fig:fig8}
\end{figure}

\noindent
In three different experiments, $p\tau_{\rm vib}$ values are reported
between 300 K and 4500 K\cite{mcn72:507,eck73:2787,bre68:4768} and
rates were calculated from $p\tau_{\rm vib}$ values and fitted to
analytical expressions.\cite{on2vrbook} Here, VR rates are calculated
for O+N$_2(v=1,j) \rightarrow$ O+N$_2(v'=0,j')$ for temperatures
between 300 K and 10000 K and are reported in Figure \ref{fig:fig8},
along with
experimental\cite{mcn72:507,eck73:2787,bre68:4768,on2vrbook} and
previous theoretical\cite{esp17:6211} results. VR rates calculated on
PESs from Ref. \citen{gam03:2545} underestimate the experiment for $T
< 9000$ K.\\

\noindent
QCT with histogram binning (QCT-HB) significantly overestimates the VR
rates for $T>800$ K while the Gaussian
binning\cite{bon97:183,bon04:106,konthesis} (QCT-GB) scheme leads to
some improvement. Since only $v=1$ to $v'=0$ transitions are
considered, determining the final quantum state is thus important and
the type of binning scheme used plays a potentially important role. It
is found that for a large number of trajectories vibrational energy
exchange was less than one quantum which leads to a state with $v' =
0$ but with high energy. Although QCT-GB partly excludes the
contributions from those trajectories and VR rates obtained from
QCT-GB are in fair agreement. It is found that considering the
trajectories with $\varepsilon_{0,j'} \leq \varepsilon_{v',j'} \leq
\varepsilon_{0,j'}+0.075$ eV (0.075 eV $\approx 0.3$ quanta) to be a
VR trajectory gives result which are in good agreement with experiment
(shown as a green line with asterisk (Modified HB, QCT-MHB). This
finding suggests that a fraction of the trajectories either do not
enter into the strong coupling region or sample it incompletely which leads
to reduced energy exchange (not fully relaxed).\\

\noindent
Yet another approach was followed in previous work\cite{esp17:6211} by
classifying the VR trajectories to `purely nonreactive' (PNR) and
`quasi-reactive' (QR) and counting the contribution only from QR
trajectories. This resulted in VR rates close to experiment for
$T>10000$ K. However, it could not improve the low temperature VR
rates. Such an analysis is also explored here. Figure S2
shows the VR rates obtained here (green line) from considering only
the QR trajectories which are close to the rates from
Ref. \citen{esp17:6211} (blue line). This points towards the
deficiencies of PESs from Ref. \citen{gam03:2545}. It is found that
the PESs from Ref. \citen{gam03:2545} describe the reactive events
quite well (reaction rates for the forward and reverse process are in
good agreement with the
experiment).\cite{gam03:2545,gam06:174303,luo17:074303,gam10:144304,esp17:6211}
Since QR trajectories are similar to reactive trajectories PESs from
Ref. \citen{gam03:2545} also results QR rates similar to present work
PESs. The PNR trajectories sample distinctly different regions in
configuration space - primarily the long range part of the PES. As the
VR rates from the present analysis agree with experiment over the
entire temperature range, the present PESs are better suited to
describe both, the low- and high-energy part of VR.\\

\noindent
These findings underline the point that different observables (here
thermal and vibrational relaxation rates) are sensitive to different
parts of the PES in the dynamics.\cite{MM.no2:2020} Reaction and
dissociation rates obtained from the earlier PESs\cite{gam03:2545} are
comparable with results from the present PESs but the vibrational
relaxation rates are different. Exploiting such a sensitivity will
allow to further improve such global PESs with the aim to provide
fully predictive simulations based on electronic structure
calculations and their accurate representation as done in the present
work.\\

\section{Conclusion}
Accurate PESs have been constructed for the lowest lying $^3$A$'$ and
$^3$A$''$ states for the [NNO] system using $>20000$
MRCI+Q/aug-cc-pVTZ energies and RKHS interpolation. Both PESs have
smaller RMSE compared with previously reported PESs in the
literature. The present PESs accurately describe all the important
interaction regions and their validity is assessed by calculating
thermal reaction rates for the N($^4S$) + NO($X^2\Pi$) $\rightarrow$
O($^3P$) + N$_2$($X^1\Sigma$) and the reverse reaction and comparing
with the available experimental results. The vibrational relaxation
rates for the O+N$_2$ collisions are also computed. Good agreement
between the experiment and theory confirms the high quality of the
present PESs and provides useful generalized Arrhenius expressions for
thermal rates. These PESs can be used to calculate state-to-state
cross sections for more coarse grained simulations such as DSMC.\\

\section{Acknowledgment}
Part of this work was supported by the United State Department of the
Air Force which is gratefully acknowledged (to MM). Support by the
Swiss National Science Foundation through grants 200021-117810,
200020-188724, and the NCCR MUST (to MM), the sciCORE computer cluster
and the University of Basel is also acknowledged.
\clearpage
\newpage

%

\end{document}


\title{Supporting Information:\\Accurate Reproducing Kernel-Based Potential Energy Surfaces for
  the Triplet Ground States of N$_2$O and Dynamics for the
  N+NO$\leftrightarrow$O+N$_2$ and N$_2$+O$\rightarrow$2N+O Reactions}

\author{Juan Carlos San Vicente Veliz}
\affiliation{Department of Chemistry, University of Basel,
Klingelbergstrasse 80, CH-4056 Basel, Switzerland}

\author{Debasish Koner} 
\affiliation{Department of Chemistry, University of Basel,
Klingelbergstrasse 80, CH-4056 Basel, Switzerland}

\author{Raymond J. Bemish} \affiliation{Air Force Research Laboratory,
  Space Vehicles Directorate, Kirtland AFB, New Mexico 87117, USA}

\author{Markus Meuwly}
\affiliation{Department of Chemistry, University of Basel,
Klingelbergstrasse 80, CH-4056 Basel, Switzerland}
\email{m.meuwly@unibas.ch}

\maketitle

\begin{figure}
\includegraphics[scale=1.8]{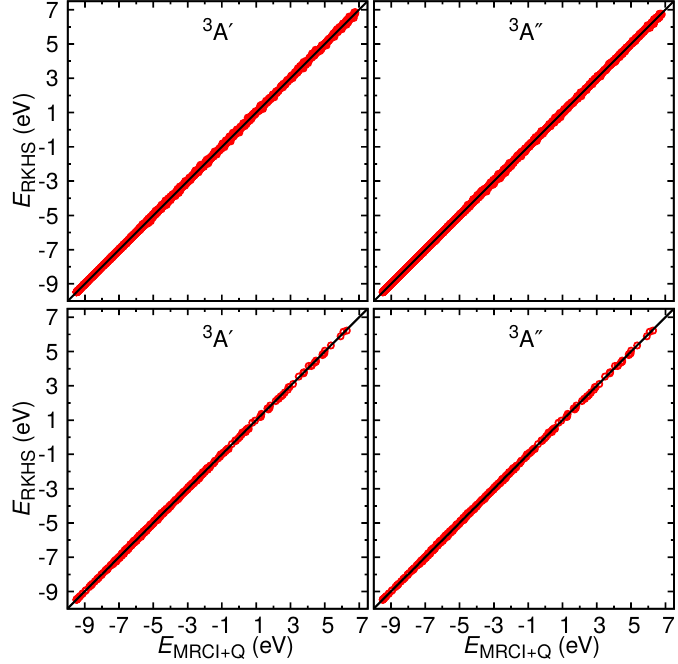}
\caption{Correlation between the {\it ab initio} and RKHS interpolated
  energies for the training grid (upper panel) and for the test grid
 (which are not part of the training grid) (lower panel) for $^3$A$'$ and $^3$A$''$ different electronic states
  of N$_2$O. Zero of energy is set to the atomization energy of N$_2$O i.e., energy
of N($^4$S)+N($^4$S)+O($^3$P). The values of $R^2$ in upper panels for the grid energies upto 6.80 eV are 0.99995 and 0.99996 for $^3$A$'$ and $^3$A$''$, respectively where as in lower panel for the test grid values of $R^2$ are 0.99985 and
0.99988 for $^3$A$'$ and $^3$A$''$, respectively.}
\label{fig:sfig1}
\end{figure}

\begin{table}[h!]
  \caption{Rate coefficients for the N($^4S$) + NO($X^2\Pi$)
  $\rightarrow$ O($^3P$) + N$_2$($X^1\Sigma$) from 100 to 20000 K
  calculated using QCT on the $^3$A$'$ and $^3$A$''$ PESs. Units are in
 cm$^3$sec$^{-1}$molecule$^{-1}$ }
	\begin{center}
	\begin{tabular}{cccc}
	\hline
	\hline
\ \ \  T (K) \ \ \ & \ \ \ \  \ \ $k^f_1(^3A')$ \ \ \ \ \ \ & \ \ \ \ \ \ $k^f_1(^3A'')$ \ \ \ \ \ \ & \ \ \ $k^f_1(^3A')$ + $k^f_1(^3A'')$ \ \ \ \\
	\hline
  100 &  0.000 &                 4.879$\times10^{-11}$ & 4.879$\times10^{-11}$ \\
  200 &  0.000 &                 3.928$\times10^{-11}$ & 3.928$\times10^{-11}$ \\
  300 &  0.000 &                 3.333$\times10^{-11}$ & 3.333$\times10^{-11}$ \\
  500 &  9.743$\times10^{-16}$ & 2.837$\times10^{-11}$ & 2.837$\times10^{-11}$ \\
  800 &  2.424$\times10^{-14}$ & 2.592$\times10^{-11}$ & 2.595$\times10^{-11}$ \\
 1000 &  7.984$\times10^{-14}$ & 2.590$\times10^{-11}$ & 2.598$\times10^{-11}$ \\
 2000 &  1.257$\times10^{-12}$ & 2.871$\times10^{-11}$ & 3.000$\times10^{-11}$ \\
 3000 &  3.759$\times10^{-12}$ & 3.234$\times10^{-11}$ & 3.610$\times10^{-11}$ \\
 3500 &  5.268$\times10^{-12}$ & 3.458$\times10^{-11}$ & 3.985$\times10^{-11}$ \\
 4000 &  7.117$\times10^{-12}$ & 3.609$\times10^{-11}$ & 4.321$\times10^{-11}$ \\
 4500 &  8.495$\times10^{-12}$ & 3.816$\times10^{-11}$ & 4.666$\times10^{-11}$ \\
 5000 &  1.077$\times10^{-11}$ & 4.018$\times10^{-11}$ & 5.095$\times10^{-11}$ \\
 5500 &  1.273$\times10^{-11}$ & 4.100$\times10^{-11}$ & 5.373$\times10^{-11}$ \\
 6000 &  1.465$\times10^{-11}$ & 4.290$\times10^{-11}$ & 5.755$\times10^{-11}$ \\
 7500 &  2.121$\times10^{-11}$ & 4.776$\times10^{-11}$ & 6.897$\times10^{-11}$ \\
10000 &  3.157$\times10^{-11}$ & 5.516$\times10^{-11}$ & 8.674$\times10^{-11}$ \\
12500 &  4.014$\times10^{-11}$ & 6.095$\times10^{-11}$ & 1.011$\times10^{-10}$ \\
15000 &  4.884$\times10^{-11}$ & 6.793$\times10^{-11}$ & 1.168$\times10^{-10}$ \\
17500 &  5.508$\times10^{-11}$ & 7.087$\times10^{-11}$ & 1.259$\times10^{-10}$ \\
20000 &  6.025$\times10^{-11}$ & 7.524$\times10^{-11}$ & 1.355$\times10^{-10}$ \\
	\hline
	\hline
  \end{tabular}
	\end{center}
\label{tab:srates1}
\end{table}

\begin{table}[h!]
  \caption{Rate coefficients for the N$'$($^4S$) + NO($X^2\Pi$)
  $\rightarrow$ N($^4S$) + N$'$O($X^2\Pi$) from 100 to 20000 K
  calculated using QCT on the $^3$A$'$ and $^3$A$''$ PESs. Units are in
 cm$^3$sec$^{-1}$molecule$^{-1}$ }
	\begin{center}
	\begin{tabular}{cccc}
	\hline
	\hline
\ \ \  T (K) \ \ \ & \ \ \ \  \ \ $k^f_2(^3A')$ \ \ \ \ \ \ & \ \ \ \ \ \ $k^f_2(^3A'')$ \ \ \ \ \ \ & \ \ \ $k^f_2(^3A')$ + $k^f_2(^3A'')$ \ \ \ \\
	\hline
100   &  0.000                & 0.000                & 0.000                 \\ 
200   &  0.000                & 0.000                & 0.000                 \\ 
300   &  0.000                & 0.000                & 0.000                 \\ 
500   &  0.000                & 0.000                & 0.000                 \\ 
800   &  0.000                & 0.000                & 0.000                 \\ 
1000  &  0.000                & 0.000                & 0.000                 \\ 
2000  &  1.881$\times10^{-16}$ & 0.000                & 1.8812$\times10^{-16}$ \\ 
3000  &  2.591$\times10^{-14}$ & 3.029$\times10^{-14}$ & 5.6202$\times10^{-14}$ \\ 
3500  &  6.513$\times10^{-14}$ & 9.275$\times10^{-14}$ & 1.5792$\times10^{-13}$ \\ 
4000  &  1.469$\times10^{-13}$ & 1.718$\times10^{-13}$ & 3.1872$\times10^{-13}$ \\ 
4500  &  2.848$\times10^{-13}$ & 3.062$\times10^{-13}$ & 5.9102$\times10^{-13}$ \\ 
5000  &  4.408$\times10^{-13}$ & 4.768$\times10^{-13}$ & 9.1752$\times10^{-13}$ \\ 
5500  &  7.233$\times10^{-13}$ & 7.878$\times10^{-13}$ & 1.5112$\times10^{-12}$ \\ 
6000  &  1.051$\times10^{-12}$ & 1.086$\times10^{-12}$ & 2.1372$\times10^{-12}$ \\ 
7500  &  2.626$\times10^{-12}$ & 2.512$\times10^{-12}$ & 5.1392$\times10^{-12}$ \\ 
10000 &  6.338$\times10^{-12}$ & 6.481$\times10^{-12}$ & 1.2822$\times10^{-11}$ \\ 
12500 &  1.187$\times10^{-11}$ & 1.117$\times10^{-11}$ & 2.3042$\times10^{-11}$ \\ 
15000 &  1.696$\times10^{-11}$ & 1.650$\times10^{-11}$ & 3.3452$\times10^{-11}$ \\ 
17500 &  2.176$\times10^{-11}$ & 2.106$\times10^{-11}$ & 4.2812$\times10^{-11}$ \\ 
20000 &  2.618$\times10^{-11}$ & 2.532$\times10^{-11}$ & 5.1502$\times10^{-11}$ \\
	\hline
	\hline
  \end{tabular}
	\end{center}
\label{tab:srates2}
\end{table}

\begin{table}[h!]
  \caption{Rate coefficients for the O($^3P$) + N$_2$($X^1\Sigma$) $\rightarrow$  N($^4S$) + NO($X^2\Pi$) from 2800 to 20000 K
  calculated using QCT on the $^3$A$'$ and $^3$A$''$ PESs. Units are in
 cm$^3$sec$^{-1}$molecule$^{-1}$ }
	\begin{center}
	\begin{tabular}{cccc}
	\hline
	\hline
\ \ \  T (K) \ \ \ & \ \ \ \  \ \ $k^b(^3A')$ \ \ \ \ \ \ & \ \ \ \ \ \ $k^b(^3A'')$ \ \ \ \ \ \ & \ \ \ $k^b(^3A')$ + $k^b(^3A'')$ \ \ \ \\
	\hline
 2800 &  0.000                 & 0.000                 & 0.000                 \\ 
 3000 &  0.000                 & 7.348$\times10^{-16}$ & 7.348$\times10^{-16}$ \\ 
 3500 &  0.000                 & 3.534$\times10^{-15}$ & 3.534$\times10^{-15}$ \\ 
 4000 &  3.330$\times10^{-15}$ & 2.079$\times10^{-14}$ & 2.412$\times10^{-14}$ \\ 
 4500 &  1.305$\times10^{-14}$ & 5.743$\times10^{-14}$ & 7.048$\times10^{-14}$ \\ 
 5000 &  2.368$\times10^{-14}$ & 1.293$\times10^{-13}$ & 1.529$\times10^{-13}$ \\ 
 5500 &  8.772$\times10^{-14}$ & 2.555$\times10^{-13}$ & 3.433$\times10^{-13}$ \\ 
 6000 &  1.575$\times10^{-13}$ & 4.374$\times10^{-13}$ & 5.950$\times10^{-13}$ \\ 
 7000 &  5.105$\times10^{-13}$ & 1.171$\times10^{-12}$ & 1.682$\times10^{-12}$ \\ 
10000 &  3.974$\times10^{-12}$ & 6.773$\times10^{-12}$ & 1.075$\times10^{-11}$ \\ 
12000 &  8.785$\times10^{-12}$ & 1.392$\times10^{-11}$ & 2.271$\times10^{-11}$ \\ 
15000 &  2.082$\times10^{-11}$ & 2.859$\times10^{-11}$ & 4.942$\times10^{-11}$ \\ 
17000 &  3.096$\times10^{-11}$ & 4.010$\times10^{-11}$ & 7.105$\times10^{-11}$ \\ 
20000 &  4.575$\times10^{-11}$ & 5.804$\times10^{-11}$ & 1.038$\times10^{-10}$ \\
	\hline
	\hline
  \end{tabular}
	\end{center}
\label{tab:srates3}
\end{table}

\begin{table}[h!]
  \caption{N$_2$ dissociation rates for the O($^3P$) + N$_2$($X^1\Sigma$) 
  collisions for 7000 to 20000 K calculated using QCT simulations on the $^3$A$'$ and $^3$A$''$ PESs. Units are in
 cm$^3$sec$^{-1}$molecule$^{-1}$}
	\begin{center}
	\begin{tabular}{cc}
	\hline
	\hline
\ \ \  T (K) \ \ \ & \ \ \ $k^d_1(^3A')$ + $k^d_1(^3A'')$ \ \ \ \\
	\hline
	7000 & 0.000 \\
	10000 & 1.061$\times10^{-13}$ \\
12000 & 5.848$\times10^{-13}$ \\
15000 & 3.915$\times10^{-12}$ \\
17000 & 6.888$\times10^{-12}$ \\
20000 & 1.720$\times10^{-11}$ \\
\hline
	\hline
  \end{tabular}
	\end{center}
\label{tab:srates4}
\end{table}

\begin{figure}
\centering
\includegraphics[scale=1.5]{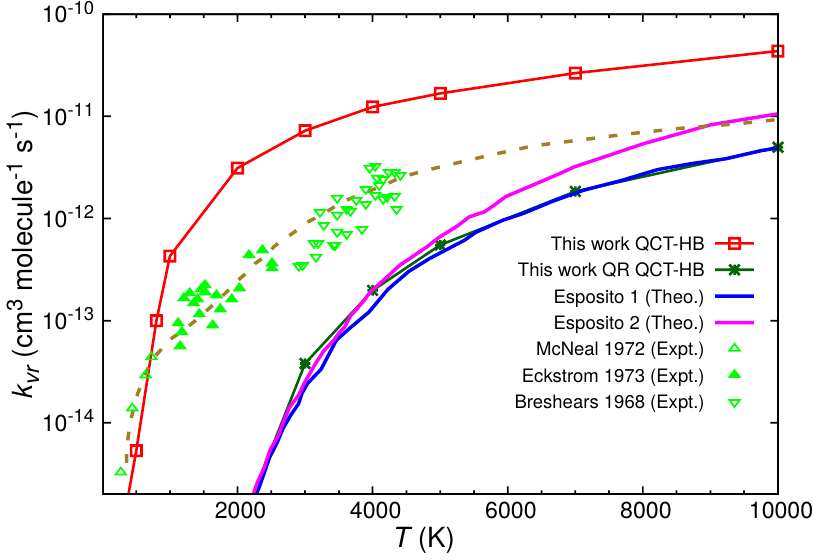}
\caption{Vibrational relaxation rates for O + N$_2$($\nu = 1$)
  $\rightarrow$ O + N$_2$($\nu' = 0$). QR QCT-HB rates calculated from
  the quasi-reactive (QR) trajectories (see text). Fit for the experimental results
  is shown as dashed olive line.}
\label{fig:sfig2}
\end{figure}